\documentstyle{mn}
\input epsf.sty
\newif\ifAMStwofonts
\AMStwofontstrue
 
\ifoldfss
  \ifCUPmtlplainloaded \else
    \NewTextAlphabet{textbfit} {cmbxti10} {}
    \NewTextAlphabet{textbfss} {cmssbx10} {}
    \NewMathAlphabet{mathbfit} {cmbxti10} {} % for math mode
    \NewMathAlphabet{mathbfss} {cmssbx10} {} %  "   "    "
  \fi
  \ifAMStwofonts
    \ifCUPmtlplainloaded \else
      \NewSymbolFont{upmath} {eurm10}
      \NewSymbolFont{AMSa} {msam10}
      \NewMathSymbol{\upi}     {0}{upmath}{19}
      \NewMathSymbol{\umu}     {0}{upmath}{16}
      \NewMathSymbol{\upartial}{0}{upmath}{40}
      \NewMathSymbol{\leqslant}{3}{AMSa}{36}
      \NewMathSymbol{\geqslant}{3}{AMSa}{3E}

       \let\ge=\geqslant
    \fi
  \fi
\fi % End of OFSS
 
\ifnfssone
  \newmathalphabet{\mathit}
  \addtoversion{normal}{\mahit}{cmr}{m}{it}
  \addtoversion{bold}{\mathit}{cmr}{bx}{it}
  \newmathalphabet{\mathbfit} % math mode version of \textbfit{..}
 \addtoversion{normal}{\mathbfit}{cmr}{bx}{it} 
  \addtoversion{bold}{\mathbfit}{cmr}{bx}{it}
  \newmathalphabet{\mathbfss} % math mode version of \textbfss{..}
  \addtoversion{normal}{\mathbfss}{cmss}{bx}{n}
  \addtoversion{bold}{\mathbfss}{cmss}{bx}{n}
  \ifAMStwofonts
    \ifCUPmtlplainloaded \else
and
your
      \UseAMStwoboldmath
      \makeatletter
      \new@mathgroup\upmath@group
      \define@mathgroup\mv@normal\upmath@group{eur}{m}{n}
      \define@mathgroup\mv@bold\upmath@group{eur}{b}{n}
      \edef\UPM{\hexnumber\upmath@group}
      \new@mathgroup\amsa@group
      \define@mathgroup\mv@normal\amsa@group{msa}{m}{n}
      \define@mathgroup\mv@bold\amsa@group{msa}{m}{n}
      \edef\AMSa{\hexnumber\amsa@group}  
      \makeatother
      \mathchardef\upi="0\UPM19
      \mathchardef\umu="0\UPM16
      \mathchardef\upartial="0\UPM40
      \mathchardef\leqslant="3\AMSa36
      \mathchardef\geqslant="3\AMSa3E

       \let\ge=\geqslant 
    \fi
  \fi
\fi % End of NFSS release 1
   
\ifnfsstwo
  \DeclareMathAlphabet{\mathbfit}{OT1}{cmr}{bx}{it}
  \SetMathAlphabet\mathbfit{bold}{OT1}{cmr}{bx}{it}
  \DeclareMathAlphabet{\mathbfss}{OT1}{cmss}{bx}{n}
  \SetMathAlphabet\mathbfss{bold}{OT1}{cmss}{bx}{n}
  \ifAMStwofonts
    \ifCUPmtlplainloaded \else
      \DeclareSymbolFont{UPM}{U}{eur}{m}{n}
      \SetSymbolFont{UPM}{bold}{U}{eur}{b}{n}
      \DeclareSymbolFont{AMSa}{U}{msa}{m}{n}
      \DeclareMathSymbol{\upi}{0}{UPM}{"19}
      \DeclareMathSymbol{\umu}{0}{UPM}{"16}
      \DeclareMathSymbol{\upartial}{0}{UPM}{"40}
      \DeclareMathSymbol{\leqslant}{3}{AMSa}{"36}
      \DeclareMathSymbol{\geqslant}{3}{AMSa}{"3E}

       \let\ge=\geqslant
    \fi
  \fi  
\fi % End of NFSS release 2
      
\ifCUPmtlplainloaded \else
  \ifAMStwofonts \else % If no AMS fonts
    \def\upi{\pi}
    \def\umu{\mu}
    \def\upartial{\partial}
  \fi
\fi

\title{Two-phase radiative/conductive equilibrium in AGN and GBH}
\author[ A. R\' o\. za\' nska, B. Czerny]
       {A. R\' o\. za\' nska$^1$, B. Czerny$^1$\\
        $^1$N. Copernicus Astronomical Center, Bartycka 18, 00-716 Warsaw, 
        Poland\\}
\begin{document}

\maketitle

\begin{abstract}

We determine simple analytical conditions for combined radiative and thermal
equilibrium between the X-ray emitting plasma and cold reprocessor 
in active galactic nuclei (AGN) and galactic black holes (GBH). These 
conditions determine the pressure at the transition
zone, which is not arbitrary in the static situation. 
The conditions for static solution are derived analytically  
for different forms of plasma heating and for cooling provided by
Compton cooling, bremsstrahlung, and 
(optionally) advective cooling. 

We conclude that if  Compton heating is the only heating mechanism,
we always achieve static equilibrium
between the phases.   For constant volume heating and radiatively cooled plasma
the static solution is 
never achieved and evaporation or condensation takes place. 
However, static solutions with no evaporation/condensation are found for 
radiative cooling supplemented with lateral advection.
Similar results are obtained for a two-temperature plasma.
In the case of a 
general prescription of
mechanical heating ($Q^{+}=h_o P^m T^{-s}$) of radiatively-cooled one-temperature
plasma, we found that static
solution is only possible, when $0 < s < 3/2$.

Such conditions apply to all models, like disc/corona solutions, ADAF flows
at their outer ADAF/disc boundary or cold clumps embedded in a hot medium. 
These models have to be reconsidered and 
supplemented either with additional condition for the pressure at the 
transition zone or with equation describing the mass exchange between the
phases.

\end{abstract}


\section{Introduction}

The broad band spectra of AGN and GBH clearly show that the accreting matter
close to a black hole consists of two phases. Relatively cool, optically 
thick phase has 
the characteristic
temperature of order of $10^5$ K in AGN and $10^7$ K in GBH, and the electron 
temperature of the hot optically thin phase is about $10^9$ K (for a review, 
see Mushotzky, Done \& Pounds 1993; Tanaka \& Lewin 1995).  

The description of the coexistence of the hot and cold medium has therefore 
to be incorporated into all models of the accretion flow, independent from 
their assumed geometry for cold and hot gas. 

Most papers on this subject model only the radiative equilibrium
within the hot and cold medium, including the radiative coupling between the
two phases (see e.g. Haardt \& Maraschi 1991, Zdziarski, Lubi\' nski \& 
Smith 1999). 
It leaves unnecessary arbitrariness in the modeling,
e.g. the value of the transition radius in ADAF based models or the fraction of
the energy liberated in the hot phase in coronal models.

However, at the boundary between the hot and cold medium there is an extreme
temperature gradient and the flow of the heat due to conduction, which is
proportional to this gradient, is essential within the boundary zone. Even
if the plasma coexist in radiative equilibrium, the conduction usually 
leads to a time evolution of the gas: either evaporation of the cold phase or
condensation of the hot phase proceeds (e.g. Begelman \& McKee 1990, McKee \& Begelman
1990). This time evolution is
specified by the efficiency of conduction. Efficient evaporation may lead
to disappearance of the innermost part of the disc, as shown by Meyer \&
Meyer-Hofmeister (1994) for the case of accretion onto a white dwarf.

The role of conduction in the context of AGN or GBH was studied only in a few
papers so far. Macio\l ek-Nied\' zwiecki, Krolik \& Zdziarski (1997) 
calculated
the vertical temperature profile for an upper part of the corona with direct 
heating of electrons
proportional to the density. R\' o\. za\' nska (1999) 
calculated the full temperature 
profile for a Compton heated corona.
Krolik (1998) and Torricelli-Ciamponi \& Courvoisier 
(1999) studied the case of a cloud model. Krolik (1998) considered the
conductive balance of the clouds with constant heating per unit
volume. Torricelli-Ciamponi \& Courvoisier (1999) calculated numerically
the temperature profile across the cold cloud/hot inter-cloud medium for a
few special cases of the heating and for cooling function including 
bremsstrahlung as well as other atomic processes important at lower 
temperature.
Finally, Mineshige et al. (1998) discussed the quiescent state of X-ray novae
taking into account the disc evaporation and  Dullemond (1999) considered the
transition zone in the case of one-temperature, advection-cooled hot corona
above an accretion disc.
  
In this paper we rediscuss the problem of condition imposed by the effect of 
conduction onto the boundary between the cold and hot phase. We show in a 
simple
analytical way which forms of heating lead to well defined condition for
pressure at the transition zone, supplementing the condition of purely 
radial equilibrium deeply within the hot layers. Heating mechanisms which
lead to heating constant per unit volume (like the $\alpha P$ scaling of
Shakura \& Sunyaev 1973), or give too steep drop of
the heating with the temperature, do not allow for any static boundary
between the two phases and in such models explicit time evolution (e.g.
evaporation rate of the cold phase) should be included.

\begin{figure}
\epsfxsize = 80 mm 
\epsfbox{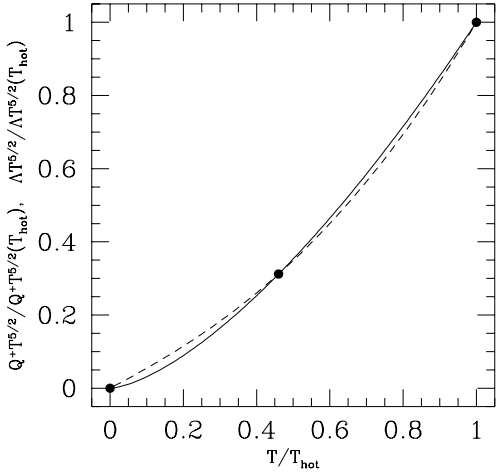}
\caption{The heating and cooling functions multiplied by the conduction flux,
normalized to their values at $T=T_{hot}$, for the case of Compton heating. 
The conductive/radiative equilibrium means that there are three crossections
between the two curves and the areas between the two curves
above and below the intermediate crossection are equal.
\label{fig:balance}}
\end{figure}

\section{General conditions of radiative/conductive equilibrium}
\label{sect:equ}

The hot plasma and cold gas may coexist both in hydrostatic and radiative
equilibrium if they are cooled by two different mechanisms 
supporting different temperature
equilibrium. However, this equilibrium in general is perturbed by the presence
of conduction and either the cold phase would evaporate or the 
hot phase would condensate
due to the heat exchange by electrons. Additional condition has to be satisfied
to sustain the two-phase equilibrium, and generally this condition can be
formulated as a condition for the pressure at the hot/cold plasma transition
zone.

The conductive heat flux $F_{cond}$ is proportional to the temperature gradient.
Therefore, it is extremely important in all those boundary regions, where the
temperature changes by three or more orders of magnitude over the small distance
determined by the Field length (Field 1965). We consider here in details the case of 
plane-parallel geometry of the transition zone but appropriate analysis can be 
repeated in other geometries (see Appendix A) 

The classical expression for $F_{cond}$ 
is given by:
\begin{equation}
F_{cond} = -\kappa_0 T^{5/2} {dT \over dz},  
\end{equation}
where the temperature $T$ changes in the direction of $z$ coordinate.
For a plasma of cosmic abundance, the conductivity constant is equal to 
(Draine \& Giuliani 1984):

\begin{equation}
  \kappa_{0} = 5.6\times 10^{-7}\phi_{c},
\end{equation}
where $\phi_{c}$ is the factor corresponding to reduction 
in the mean free path due to
magnetic fields and turbulence ($\phi_{c}=1$ 
for equal ion and electron temperature).

This classical expression has to be replaced with the expression for saturated
conduction flux if the mean free path of electrons becomes larger than the
characteristic scale high for the temperature. However, in the context of AGN
or GBH this is not  necessary (e.g. R\' o\. za\' nska 1999).

The energy balance across the boundary between the hot and cold medium in the
static case is given by:
\begin{equation}
{dF_{cond} \over dz} = Q^+ - \Lambda,
\label{eq:gener}
\end{equation}  
where $Q^+$ determines the (radiative or non-radiative) heating of the gas and
$\Lambda$ describes the radiative cooling. The boundary conditions are set by
the requirement that the conductive flux vanishes both
deeply in the cold zone and deeply in the hot zone:
\begin{equation}
F_{cond}(z_{cold}) = 0 = F_{cond}(z_{hot}),
\label{eq:surflux}
\end{equation}
and the hot zone remains in radiative equilibrium:
\begin{equation}
Q^+(z_{hot})=\Lambda(z_{hot}).
\label{eq:rad}
\end{equation}

Equation (\ref{eq:gener}) with boundary conditions (\ref{eq:surflux}) 
and \ref{eq:rad} secure that neither evaporation nor condensation take place
and the boundary between the two phases remains at its equilibrium position.

Condition (\ref{eq:surflux}) can be equivalently expressed in the integral form:
\begin{equation}
\int_{z_{cold}}^{z_{hot}} (Q^+ - \Lambda)dz = 0.
\end{equation}

Those conditions can be implemented numerically into the complete set of equations
determining the transition zone between the hot and cold medium. However, we can
obtain simplified analytical conditions determining the equilibrium transition 
zone.

In most cases studied in detail in the context of stratification in AGN or GBH  
the rapid change of the temperature (and density) 
by a few orders of magnitude happens practically under
the condition of almost constant pressure, as noticed by Torricelli-Ciamponi
\& Courvoisier (1998) and Krolik (1998), and used
in their approach.  

The condition of constant pressure allows to relate the density of the gas to its
temperature:
\begin{equation}
\rho = { P \mu m_H \over T k}.
\end{equation}
with the value of molecular weight $\mu = 0.5$ for cosmic
chemical composition. Boltzmann constant is denoted as $k$ and the mass of hydrogen atom
as $m_H$.

This gives the opportunity to express all quantities as functions of the temperature
if the heating term also depends entirely on the density, temperature and 
constant factors not varying across the boundary region. Equation 
(\ref{eq:gener})
can be solved in general form by the change of variables from $z$ to $T$
(and from $d/dz$ to $ F_{cond} T^{-5/2}/\kappa_o d/dT$): 
\begin{equation}
F_{cond}^2(T_{cold})-F_{cond}^2(T_{hot}) = 2 \int_{T_{cold}}^{T_{hot}} 
(Q^+ - \Lambda) \kappa_0 T^{5/2} dT.
\end{equation}

This formal solution can be directly used to formulate the conditions of 
radiative/conductive equilibrium. We further simplify the conditions noticing
after Krolik (1998) that the cold medium temperature is much lower than the
hot medium temperature, so it can be approximated by zero. Imposing boundary 
conditions (\ref{eq:surflux}) and (\ref{eq:rad}) we then finally arrive 
at two conditions:
\begin{equation}
\int_0^{T_{hot}} (Q^+ - \Lambda) \kappa_0 T^{5/2} dT=0,
\label{eq:fin1}
\end{equation}
and:
\begin{equation}
Q^+(T_{hot})=\Lambda(T_{hot}).
\label{eq:fin2}
\end{equation}

Those two conditions determine both the hot medium temperature and the pressure
at the transition zone if the heating prescription is assumed and the cooling
mechanisms specified.

\section{Special solutions}

The cooling mechanisms acting in AGN and GBH are relatively well known. Here
we consider Compton cooling and bremsstrahlung:
\begin{equation}
\Lambda(\rho, T) = F_{rad} \kappa_{es} {4k  \over m_e c^2}\rho T + B \rho^2 T^{1/2},
\end{equation}
and expressing the density $\rho$ by the pressure $P$ constant through the
transition zone we have:
\begin{equation}
\Lambda(\rho, T) = F_{rad} \kappa_{es}  \frac{4 \mu m_{H}}{m_e c^2} P + 
B ({\mu m_H \over k})^2 P^2 T^{-3/2},
\label{eq:lam}
\end{equation}
where $\kappa_{es}$ is opacity coefficient for electron scattering , $m_e$ -  mass 
of electron, $c$ - velocity of light, and $B$ - bremsstrahlung cooling constant, which
equals $B=6.6 \times 10^{20} {\rm erg}/{\rm s cm}^2 {\rm g}^2$.
Since we approximated the cold temperature by zero we do not have to include
complex bound-free and line cooling important at the lowest temperatures.
As was noticed by Torricelli-Ciamponi \& Courvoisier (1998), the detailed 
description
of the processes at lower temperatures is not essential since the conductive
flux is proportional to $T^{5/2}$ and the integrand is strongly 
weighted towards $T_{hot}$.

We also neglect here the synchrotron cooling. However, if the magnetic field 
in the hot plasma is important the synchrotron cooling term should be included.
We discuss the special case of advection cooling in 
Section~\ref{sub:advec}.

As for heating, the physical process leading to observed high temperature of 
electrons is not known. Several parameterization of this process are used
in modeling. Therefore we study here the most frequently used assumptions
considering both one-temperature and two-temperature plasma. In this
last case we have to supplement the equations with electron/ion Coulomb
coupling.

\subsection{Compton heating}
\label{sec:den}

Compton heated plasma typically achieves temperature of order of
$10^7 - 10^8$ K, depending on the spectral shape of the incident radiation.
This is not enough for X-ray source in AGNs and GBHs (see introduction), but 
Compton heated plasma is often considered,e.g. as an outer corona in AGN or
as confining medium for broad line clouds in AGN, so it is interesting
to check how equations  \ref{eq:fin1} and \ref{eq:fin2} work in this 
simplest example.

The term describing Compton heating can be written in the usual form:
\begin{equation}
Q^+ = F_{rad} \kappa_{es} {4 k \over m_e c^2} \rho T_{IC},
\end{equation}
where $T_{IC}$ is the Inverse Compton temperature.

Equations (\ref{eq:fin1}) and (\ref{eq:fin2}) lead to the conditions:
\begin{equation}
T_{hot}={7 \over 15}T_{IC},
\end{equation}
and 
\begin{equation}
P = {8 \over 15}F_{rad} \kappa_{es}{4 k^2 \over B \mu  m_H m_e c^2} T_{IC}^{3/2}.
\end{equation}

These conditions (see also Figure~\ref{fig:balance}) reproduce well the 
requirements for the coexistence of the
hot and cold phase based purely on radiative stability, although the 
coefficients are slightly different. We can introduce the ionization parameter 
$\Xi$ of Krolik, 
McKee \& Tarter (1981):
\begin{equation}
\Xi = {F_{rad} \over cP}.
\end{equation}
The requirement for the pressure 
leads to the same scaling 
with the Inverse Compton temperature 
\begin{equation}
\Xi^* \propto T_{IC}^{-3/2}
\end{equation}
as determined
by Begelman, McKee \& Shields (1983).

\subsection{Constant volume heating and $\alpha$ P heating}
\label{sec:con}

Looking for additional heating mechanism, which can increase the temperature
of an  X-ray source up to the 
$10^9$K we consider $\alpha P$ heating of the hot plasma (Shakura \& Sunyaev
(1973), Witt, Czerny \&  \. Zycki (1997), Krolik (1998)). 
Since we assume that the transition zone is narrow enough to be approximated
by $P = const$ both heating parameterizations are equivalent.

Denoting the constant heating as $Q^+_{const}$, conditions (\ref{eq:fin1}) 
and (\ref{eq:fin2}) are satisfied only if:
\begin{equation}
P={m_e c^2 \over 4 \mu m_H} {Q^+_{const} \over F_{rad} \kappa_{es}}, 
\label{eq:vol}
\end{equation}
and
\begin{equation}
T_{hot}=\infty, \rho_{hot}=0.
\end{equation}

Finite required value of the pressure is accompanied by condition that 
bremsstrahlung
cooling is negligible. It effectively means that any finite temperature
will lead to the violation of one of the two conditions. Radiative 
equilibrium far from the cold medium would lead to unbalanced conductive flux
and subsequent evaporation of the cold medium. Static solutions are never 
achieved. 

If only the integral condition (i.e. \ref{eq:fin1}) is used we reproduce the
results expressed by Krolik (1998) as his $H_{net} = 7/4$ (his equation
6) or by Witt, Czerny \& \. Zycki (1997) as $\Xi \propto T_{hot}^{-3/2}$.
Such a condition for pressure minimizes the effect of evaporation, therefore
its use is not entirely meaningless, but it does not provide strict static
equilibrium. 

Similar results were obtained numerically for the case of spherically symmetric
cloud by Torricelli-Ciamponi \& Courvoisier (1998). They used better 
description of the cooling, extending it down to the temperatures when 
other atomic processes are more efficient than bremsstrahlung. 
Their solutions derived
for a number of heating functions satisfied the integral condition but 
displayed strong second derivative of the temperature, $d^2T/dz^2 \propto
Q^+ - \Lambda$ (see their Appendix A). 

\subsection{Constant volume heating with radiative/advective cooling}
\label{sub:advec}

The models with radial advection and vertical stratification of the hot/cold
accreting gas allow to parameterize the advection cooling term as
\begin{equation}
{Q_{adv} \over Q^+} = \delta {T \over T_{vir}},
\end{equation}
with dimensionless coefficient $\delta$ replacing the radial derivatives of
thermodynamical parameters of the flow and (here constant) $T_{vir}$ being the 
local virial 
temperature (see e.g. Abramowicz et al. 1995, Janiuk, \. Zycki \& Czerny 1999).

Both equations (\ref{eq:fin1}) and (\ref{eq:fin2}) can now be satisfied. 
The condition for the temperature can be written in the implicit form
\begin{equation}
\delta {T_{hot} \over T_{vir}} = {27 \over 35}{Q^+ - \Lambda_C \over Q^+},
\end{equation}
where $\Lambda_C$ is the Compton cooling term (i.e. first term of 
Equation~\ref{eq:lam}) proportional to the pressure $P$. 
The value of the pressure is given by
\begin{equation}
P^2 = {8 k^2(Q^+ - \Lambda_C)T_{hot}^{3/2} \over 35 \mu^2 m_H^2 B}.
\end{equation}
If the Compton cooling is negligible, those conditions can be used directly.
Otherwise, explicit solutions for $T_{hot}$ and $P$ have to be 
derived numerically.
Evaporation of the disc underlying the advective flow is not expected.
 
Dullemond in his paper (1990) considered hot corona on the top of
the  cold disc
with exactly the same heating and cooling processes like in this section,
but for arbitrary value of the pressure expressed by his dimensionless 
constant $C$.
He concluded that disc has to evaporate, because the heat flux deep inside
cold phase stays non zero (Eq. 32 of Dullemond 1990). 
His calculations were  done for $C=0.15$, but we can find that for $C=0.1548$
both conditions are satisfied: energy balance equation and 
zero heat flux inside the disc, so the evaporation is not required.
 
\subsection{General mechanical heating and radiative cooling}

Mechanical heating of the hot phase can in most cases be parameterized in the
general form as:
\begin{equation}
Q^+= h_o P^{m}T^{-s},
\label{eq:genheat}
\end{equation}
(see Torricelli-Ciamponi \& Courvoisier 1998 and the references therein).

Conditions  (\ref{eq:fin1}) and (\ref{eq:fin2}) reduce to:
\begin{equation}
T_{hot}^s={7 h_o m_e c^2 \over 12 \mu m_H} {P^{m -1} \over  F_{rad} \kappa_{es}} 
{3/2 - s \over 7/2 - s},
\end{equation}
and
\begin{equation}
P ^{m - 2} = {3 B \mu^2 m_H^2 \over 4 h_o k^2 }{7/2 - s \over s} T_{hot}^{s - {3 \over 2}}.
\end{equation}
Combining those two equations we find prescription for pressure:
\begin{eqnarray}
P ^{[{3 \over 2s} (m-1)-1]} & = &\frac{3B \mu^2 m_H^2}{4 h_o k^2} 
\left(\frac {7 h_o m_e c^2}{12 \mu m_H F_{rad} \kappa_{es}}\right)^{(1-{3\over 2s})} 
\times \nonumber \\
 & &\frac {3/2-s}{s} \left(\frac{7/2-s}{3/2-s}\right)^{ 3\over 2s }
\end{eqnarray}

All equations above can be satisfied, leading to positive values for both the
pressure $P$ and the hot plasma temperature, $T_{hot}$ only if
\begin{equation}
 0 < s < {3 \over 2}
\label{eq:conds}
\end{equation}

We see that the problem met by Torricelli-Ciamponi \& Courvoisier (1998)
to satisfy both the integral condition across the cold/hot boundary 
{\bf and} the radiative equilibrium in the hot phase arose because of
unfortunate choice of the index $s$ (equal 1.5 and 2.25 in their case,
see also Appendix A).

\subsection{Two-temperature plasma}
\label{sect:two}

Efficient predominant heating of ions may lead to a two-temperature plasma, 
with electron temperature $T_e$ lower than ion temperature, $T_i$. The heat 
exchange of
plasma is provided by the Coulomb coupling and possibly other mechanisms like
plasma waves. For simplicity, we can assume 
exclusive heating of ions, Coulomb coupling as the only heat transfer 
mechanism, and we can neglect the contribution to the pressure from the
electrons.

For any ion heating parameterization $Q^+(T_i)$ we have following heating and
cooling functions for electrons:
\begin{equation}  
 Q^+ = \frac {3}{2} \frac {k}{m_H} D \rho^2 (T_i-T_e)T_e^{-3/2},
\end{equation} 
\begin{equation}
\Lambda(\rho, T_e) = F_{rad} \kappa_{es} \rho {4k T_e \over m_e c^2} +
+ \delta Q^+ {T_i \over T_{vir}}+ B \rho^2 T_e^{1/2},
\end{equation}
where $D$ is Coulomb coupling constant equal $D=2.44 \times 10^{21} 
{\rm ln} {\Lambda}$ and  ${\rm ln} {\Lambda} \approx 20$.
The density can be determined by the pressure:
\begin{equation}
\rho= {P m_H\over k T_i}, 
\end{equation}
which allows to express all quantities as functions of $T_e$. The equilibrium
conditions are given by Equations (\ref{eq:fin1}) and (\ref{eq:fin2}), as 
before, only for
the electron temperature. These equations have to be solved numerically, after specifying $Q^+(T_i)$ 

We considered in detail a few special cases. 

For $Q+ = const$, without advection,
the result is the same
as in the case of one-temperature plasma (see Section \ref{sec:con}). 
If radiation
balance is satisfied the integral condition is never satisfied for a finite
hot medium temperature. Strictly static solutions do
not exists.

In the case of $Q^+ \propto \rho$, without advection, the solution was found for
a representative set of parameters, again similarly to the case of one-temperature
plasma (see Section \ref{sec:den}. Solutions could also be found for 
$Q^+ = const$, but with the advection term included.

\section{Discussion}
\label{sect:diss}

Static solution for two-phase 
equilibrium in AGN and GBH with radiative/conductive energy exchange 
between phases may, or may not exist, depending on assumed heating 
and cooling mechanisms operating in the hot phase and in the transition zone. 

In the case of the Compton heated corona (R\' o\. za\' nska 1999)
we always obtain static solution, and the required pressure depends on Inverse 
Compton temperature.

Compton heating, however, is not sufficient to heat up the X-ray emitting
plasma in
those objects and other mechanisms must operate, extracting the gravitational
energy of the accreting material. The exact path of the energy flow is
not known so we considered typical heating parameterizations.

We found that the evaporation/condensation equilibrium (i.e. static solution)
for radiatively cooled plasma is possible if the heating is a moderately 
decreasing function of the temperature (see Eqs.~\ref{eq:genheat} and 
\ref{eq:conds}). 

Frequently used assumption that the heating is proportional to $\alpha P$ 
(Shakura \& Sunyaev 1973) does
not lead to equilibrium for a radiatively cooled plasma. Accretion disc 
covered by such a corona would evaporate continuously. However, if coronal 
advection is allowed, equilibrium may be restored.

Detailed computations of such an equilibrium, particularly for a 
two-temperature plasma, have to be performed numerically 
(see Section~\ref{sect:two}), so the existence of solutions for any 
particular set of parameters should be checked.
A number of very interesting models of the accretion flow have been recently
proposed, e.g. Esin, McClintock \& Narayan (1997), Collin-Souffrin et al. 
(1996), Poutanen, Krolik \& Ryde (1997), Krolik (1998), 
Nayakshin, Rappaport \& Melia (1999), 
and we can apply 
the approach outlined
in the present paper to these models.
Our analysis is not applicable when coronal heating occurs in small compact
regions, like in case of magnetic field reconnection. 

In realistic two-phase flow we probably have two mechanisms leading to the mass
 exchange between the hot and the cold phases. First mechanism should be 
directly related to the dynamics of the flow, like heating the plasma in magnetic 
loops or instabilities within the cold disc, leading to disruption of the cold 
flow and hot optically thin plasma production. This mechanism would be 
responsible for the appearance of the hot plasma in the first place. The 
second mechanism operates when the hot plasma already exists: 
evaporation/condensation due to conduction leads to further mass exchange. 
This second mechanism was the subject of the present paper and we formulated 
the conditions under which no further mass exchange proceeds.

However, in real situation the study of the appropriate time scales based on
adopted dynamical model should be done. 
As an example, we consider a scenario with energy in both phases  released 
via viscosity dissipation (situation calculated in Section \ref{sec:con}). We
can estimate accreting and isobaric cooling time scale
defined as follows:
\begin{equation}
t_{acc}=\frac{r}{V_r},
\end{equation}
\begin{equation}
t_{cool}=\frac{5}{2} \frac{p}{n^2 \Lambda},
\end{equation}
when $r$ is the distance from the central object, $V_r$ - radial velocity.
To compare those two time scales we assume $n^2 \Lambda$ to be Compton 
cooling
and we use Eq. \ref{eq:vol} to determine constant volume heating.
The ratio of both time scales can be expressed by Keplerian velocity and
the velocity of sound as:
\begin{equation}
\frac{t_{acc}}{t_{cool}}=\frac {3}{5} \frac{V_K^2}{V_s^2}=\frac {3}{5}
\frac{T_{vir}^2}{T^2}.
\end{equation}
For accreting corona, if the 
temperature is much smaller than the virial temperature, 
this ratio is always much
bigger than $1$ and radiative/conductive equilibrium should be achieved.
Since in the case of constant heating without advection we did not find a
static solution it means that rapid evaporation or condensation will result
in transition to be an one phase medium.
However, if the corona temperature is close to the virial temperature, the 
evaporation time scale is comparable to the viscous time scale, so the 
two phase medium may exist but continuous evaporation of the cold material
should be calculated and included into the model.

Similar discussion of the thermal and viscous time scale has to be repeated
for other two-phase accretion models in order to check whether the initially 
non-equilibrium conditions will be rapidly changed into practically static 
solutions by adjusting the pressure and the temperature profile through 
evaporation/condensation and radiative heating/cooling. 

\label{sect:conc}

\section*{Acknowledgments}

We thank Piotr \. Zycki for helpful discussion and comments
and the anonymous referee for emphasizing the importance of discussion
of spherical clouds. 
This work was supported in part 
by grant 2P03D01816 of the Polish State Committee for 
Scientific Research.

\appendix

\section{Application for the spherically symmetric system}
\label{app:sfer}

Similar analysis can be repeated for a spherically symmetric, dense 
cold cloud embedded in the hot, optically thin plasma, as it was done
by McKee \& Begelman (1990) and
Torricelli-Ciamponi \& Courvoisier (1999).
In such  case the existence of static solution 
strongly depends not only on the form of the heating function,
but also on the size of a cloud, $R_{cl}$.  

For spherically symmetric case the energy
balance across the boundary between hot and cold medium in the static case is
given by:
\begin{equation}
 {1 \over r^2} {d \over dr} (r^2 F_{cond}) = Q^+ - \Lambda, 
\end{equation}
where $r$  is the radial coordinate equaling zero at the cloud center.
Adopting classical expression for $F_{cond}=-\kappa_0 T^{5/2} dT/dr$
and assuming the same boundary conditions as in plane-parallel case,
the equilibrium between the cloud and the surrounding medium is 
described by:
\begin{equation}
\int_{T_{cold}}^{T_{hot}} r^4 (Q^+ - \Lambda) \kappa_0 T^{5/2} dT=0.
\label{ap:war}    
\end{equation}
Now, $T_{cold}$ is the temperature of the cold cloud.
Static solution also requires that far from the transition 
interface the radiative equilibrium is achieved. 

Those two conditions: Eq.\ref{ap:war} and Eq. \ref{eq:fin2} cannot be solved in
general since the first condition requires the knowledge of
$T(r)$ explicitly in the spherical geometry.
Nevertheless, we can analyze the situation qualitatively, studying asymptotic behavior
of clouds with radii very small or very large in comparison to
the Field length $ \lambda_{F}$, defined as the characteristic maximum length 
over which the energy exchange by thermal conduction is negligible and
radiative processes are dominated (Field 1956). 

It was shown by McKee \& Begelman (1990) for the case of radiatively heated clouds,
and by Torricelli-Ciamponi \& Courvoisier (1999) 
for more general heating functions,
that when $R_{cl} << \lambda_{F}$ the static solution for spherically symmetric
cloud is never achieved, such a small cloud should always evaporate.
But McKee \& Begelman (1990) found that for $R_{cl} \ge \lambda_{F}$ 
there is the value of pressure at which the mass exchange 
is stopped and static 
solution condition can be satisfied. For higher pressure a cloud will condense 
(condensation occurs as a negative sign of evaporation rate).
The range of pressures for which a cloud can condense is wider for 
cloud radii much larger than the Field length.

Those results in the limit of small cloud radii $R_{cl}/\lambda_{F}<< 1 $ were 
confirmed analytically by Torricelli-Ciamponi \& Courvoisier (1999) who made appropriate 
Taylor expansion and showed that static solution is never achieved.

We made a similar Taylor expansion in the opposite limit, for $\lambda_{F}/R_{cl}<< 1 $.
For such purpose the expansion of $r$ is done around point $T1$ which is the 
second crossection of cooling and heating functions 
(see Fig. \ref{fig:balance}).
For the value of temperature $T1$ the radius is denoted as 
$R_{cl}$, so the expansion is given by:
\begin{eqnarray}
 r(T) \,\,\, = & R_{cl}+{dr \over dT}(T-T1) \, +... \nonumber \\
 r^4(T)  = & R_{cl}^4 + 4R_{cl}^3{dr \over dT}(T-T1)... 
\end{eqnarray}
Substituting the expansion to the Eq. \ref{ap:war} we obtain:
\begin{eqnarray}
\lefteqn {\int_{T_{cold}}^{T_{hot}} (Q^+ - \Lambda)\kappa_0 T^{5/2} dT  +} 
\nonumber \\
& + & {4 \over R_{cl}} {dr \over dT} \int_{T_{cold}}^{T_{hot}} 
(Q^+ - \Lambda) (T-T1) \kappa_0 T^{5/2}  dT=0.
\label{ap:war3}    
\end{eqnarray}

The second term is always positive so the conditions derived for a plane
parallel geometry are replaced by inequalities in the case of a 
spherical geometry. Quantitative estimate of this correction can be done 
using the concept of the Field length. We show an example of the result for
the case of mechanical heating $m=1$ and
$s=1$ (see Sec. 3).
Since the Field length is inversely proportional to the characteristic 
temperature scalelength, the 
derivative  $d ln r /d ln T = \lambda_{F}/R_{cl}$.
We estimate that the second integral in Eq. \ref{ap:war3} is of the order 
of $Q^{+}\kappa_0 T^{7/2}$. Now this condition   
together with fact that for $T_{hot}$ the radiative 
equilibrium is achieved, allow to determine the temperature of the 
hot phase and
the pressure of the transition. Comparing to the temperature 
and pressure in case of the 
plane parallel geometry when $R_{cl} = \infty$, we obtain:
\begin{equation}
T_{hot}(R_{cl})= T_{hot}(R_{cl} = \infty)(1-10 {\lambda_{F} \over R_{cl}}),
\end{equation} 
and
\begin{equation}
P(R_{cl}) = P (R_{cl} = \infty)(1+ {15 \over 4} {\lambda_{F} \over R_{cl}}).
\end{equation} 
In the limit of $\lambda_{F}/R_{cl}<< 1 $ for spherical clouds 
the static solution is achieved, if the solution exists in the case of a 
plane parallel geometry.

Between those two limits the situation is more complicated.
To find the value of pressure together with spherical
cloud radius for which the static solution 
is satisfied the proper numerical computations should be done and
we will address this problem in our future work. 
  
\ \\
This paper has been processed by the authors using the Blackwell
Scientific Publications \LaTeX\  style file.

\end{document}